\documentclass[twocolumn,aps,showpacs,prl]{revtex4}
\usepackage{graphicx}
\usepackage{epstopdf}
\usepackage{amsmath}
\usepackage{multirow}
\begin{document}

\title{Effect of the Wyckoff position of K atom on the crystal structure and electronic properties of KFe$_2$Se$_2$ compound}

\author{Xun-Wang Yan$^{1,3}$}
\email{xwyan@itp.ac.cn}
\author{Miao Gao$^{2}$}

\date{\today}

\affiliation{$^{1}$School of physics and electrical engineering,
 Anyang Normal University, Henan 455000, China }

\affiliation{$^{2}$Department of Physics, Renmin University of
China, Beijing 100872, China}

\affiliation{$^{3}$Institute of Theoretical Physics, Chinese Academy of Sciences, Beijing 100190, China }

\begin{abstract}
By the first-principle electronic structure calculations, we study the effect of Wyckoff position of K atom on the crystal and electronic structures of KFe$_2$Se$_2$ compound.
When the K atoms take up the Wyckoff position $2a$, $2b$ or $4c$ (the related structure of KFe$_2$Se$_2$ is called as Struc-2a, Struc-2b or Struc-4c), the calculated lattice constant $c$ is in the range of 13.5$\sim$14.5 \AA{}, 15.5$\sim$16.7 \AA{} or 18.6$\sim$19.1 \AA{}.
Three concentric cylinder-like Fermi surfaces emerge around $\Gamma$-Z in Brillouin Zone for the Struc-2b in nonmagnetic state, different from the Struc-2a and Struc-4c.
The Fe-Se-Fe angle is 107.8$^\circ$, 108.8$^\circ$ or 110.7$^\circ$ in the collinear anti-ferromagnetic state and the superexchange interaction $J_2$ between two next neighbor Fe moments is 13.08 meV/S$^2$, 20.75 meV/S$^2$ or 11.86 meV/S$^2$ in Struct-2a, Struc-2b or Struc-4c structure respectively.
The Struc-2b and Struc-4c have a good correspondence to the newly discovered superconducting phases with T$_c$=40 K and T$_c$=30 K in KFe$_2$Se$_2$.
Our findings provide a reasonable approach to understand the existence of multiple superconducting phases in alkali metal intercalated FeSe superconductor.

\end{abstract}


\maketitle

\section{Introduction}
The recent observation of superconductivity about 30 K in the layered iron selenides compound K$_{0.8}$Fe$_2$Se$_2$ \cite{chen} initiated extensive research on the iron selenide superconductors, which was formed by intercalating alkali metal between FeSe layers and had the ThCr$_{2}$Si$_{2}$-type structure (Fig. 1(a)).
In contrast to the iron arsenide superconductors, A$_y$Fe$_{2-x}$Se$_2$ (A = K, Rb, Cs or Tl) is a new class of iron based superconductor. Their electronic structures are very different from those in the iron arsenide compounds, that is, only electron Fermi surfaces are presented around the zone corners while no hole Fermi surface near the zone center,
which is confirmed by the experimental measurement \cite{feng-naturem} and the previous first principle calculations \cite{yan-KFeSe}.

The crystal structures of A$_y$Fe$_{2-x}$Se$_2$ are complex since for different $x$ values there are different Fe vacancy arrangement schemes.The X-rays, transmission electron microscopy, and neutron scattering measurements indicate that the composition of K-intercalated FeSe superconductors was close to K$_y$Fe$_{1.6}$Se$_2$ with a fivefold expansion of the parent ThCr$_2$Si$_2$ unit cell in the $ab$ plane, namely a $\sqrt{5}\times \sqrt{5}$ Fe-vacancies superstructure \cite{basca,bao1}.
Another experiment show that there exists a rhombus Fe vacancy pattern related to K$_y$Fe$_{1.5}$Se$_2$ \cite{fang}.
However it is reported that in the same sample there exist two distinct phases, the insulating phase with well-defined $\sqrt{5}\times \sqrt{5}$ Fe-vacancies superstructure and another KFe$_2$Se$_2$ phase containing no Fe vacancies, the latter being suggested to be responsible for superconductivity \cite{weili}. The composition and structure responsible for the suerconductivity is still in debate.

Most recently, Li, Na, Ca, Ba, Yb, or Eu layer is inserted between adjacent FeSe layers to optimize the superconduting properties of the iron selenides compounds
\cite{chenxiaolong1}.
It is the surprising phenomenon that two superconducting phases were observed in K$_{0.8}$Fe$_{2}$Se$_2$ with the one T$_c$ of 30 K and another T$_c$ of 40 K.
In addition, two kinds of crystal unit cell with unexpectedly long lattice constant $c$=16.14 \AA{} and $c$=20.48 \AA{} were found to correspond to two superconducting phases, and the $c$=20.48 \AA{} unit cell disappeared after one hour's exposure to air accompanying the T$_c$ change from 40 K to 30 K.
In the other reports, two superconducting phases were ever observed in K$_{x}$Fe$_{2-y}$Se$_2$ \cite{chengenfu, sunliling, zhangqingming}.
However, there are rare theoretical studies on the difference of crystal and electronic structures between these superconducting phases, and the reason for higher Tc in new superconducting phase.

The Wyckoff position of K atoms has a great influence on $c$ axis length of crystal unit cell and the Fermi surface characters in KFe$_2$Se$_2$. In experiment the occupations of different Wyckoff positions by alkali metal atoms are observed \cite{NH3}.
%So the different Wyckoff position occupied by K atoms can be considered as the most possible factor to result in the three superconducting phases with different lattice parameter $c$. The fact that no clear peak due to N-H vibrations was detected in infrared spectroscopy measurements \cite{chenxiaolong1} exclude the entrance of NH$_3$ into KFe$_2$Se$_2$, which is thought as another impossible reason to produce the multiple phases of KFe$_2$Se$_2$.
Meanwhile, the fact that no clear peak due to N-H vibrations is detected in infrared spectroscopy measurements \cite{chenxiaolong1} excludes the existence of NH$_3$ in KFe$_2$Se$_2$ synthesized by the liquid ammonia method. So we can deduce that three superconducting phases of KFe$_2$Se$_2$ with different lattice parameter $c$ should correspond to the different Wyckoff position occupied by K atoms.
In the paper, we focus on the effect of different Wyckoff position of K atom in space group I4/mmm on the crystal structure and electronic properties of KFe$_2$Se$_2$, with K atom sited in $2a$, $2b$ or $4c$ Wyckoff position  respectively. The paper was organized as follows.
Firstly, we optimize the crystal structure and obtain the reasonable lattice constants of KFe$_2$Se$_2$, in agreement with the experimental values of $c$=14.04 \AA{}, $c$=16.16 \AA{}, and $c$=20.48 \AA{} in three superconducting phases.
Secondly, the electronic structures of Struc-4c is similar to those of Struc-2a implying they have the similar superconduting behavior. The Fermi surface of Struc-2b exhibits the different characters, that is, the concentric cylinder like Fermi surfaces are presented around $\Gamma$-Z, which should have a necessary connection to the superconductivity with the higher T$c$=40 K.
Lastly, from three aspects of the Fe-Se-Fe angle, electronic structures and the next neighbor superexchange interaction $J_2$, we demonstrate that the Struc-2b structure is closely related to the superconducting phase with T$_c$=40 K, while Struc-4c to the superconducting phase with T$_c$=30 K in KFe$_2$Se$_2$ synthesized by the ammonothermal method.

\section{Method and details}
In our calculations the plane wave basis method was used \cite{pwscf}. We adopted the generalized gradient approximation
(GGA) with Perdew-Burke-Ernzerhof formula \cite{pbe} for the exchange-correlation potentials. The ultrasoft pseudopotentials
\cite{vanderbilt} were used to model the electron-ion interactions. After the full convergency test, the kinetic energy cut-off and the charge density cut-off of the plane wave basis were chosen to be 800 eV and 6400 eV, respectively. The Gaussian broadening technique was
used and a mesh of $16\times 16\times 12$ k-points were sampled for the Brillouin-zone integration. In the calculations, the lattice parameters with the internal atomic coordinates were optimized by the energy minimization.
The Struc-4c tetragonal unit cell should have included four K atoms, but only two K atoms is included in order to keep the same composition K:Fe:Se=1:2:2 for Struc-2a, Struc-2b and Struc-4c.

\begin{table} [b]
\caption{\label{tab:table1}
The lattice parameters $c$, the Fe-Se-Fe bond angle $\alpha$ and the Se height $H_{Se}$ to Fe plane of KFe$_2$Se$_2$ in several magnetic orderings for the Struc-2a, Struc-2b and Struc-4c phases. In Bicoll-AFM order the lattice constants $a$ shrinking lead to the decrease of $\alpha$ along the FM direction and the distortion of Fe layer lead to the fluctuation of $H_{Se}$ , shown in bracket. For the Struc-2b the Bicoll-AFM state is not stable.}
\begin{ruledtabular}
\begin{tabular}{llccc}
%\textrm{Left\footnote{Note a.}}&
%\textrm{Centered\footnote{Note b.}}&
%\multicolumn{1}{c}{\textrm{Decimal}}&
%\textrm{Right}\\
structure &magnetic phase&$c$(\AA) &$H_{Se}(\AA)$&$\alpha$($^\circ$)\\
\colrule
          &Nonmag   & 13.30 & 1.29 & 112.3\\
Struc-2a   &FM       & 14.04 & 1.62 & 98.7\\
          &Neel-AFM & 13.67 & 1.37 & 109.5\\
          &Coll-AFM & 13.79 & 1.42 & 107.8\\
          &Bicoll-AFM & 14.45 &1.45&107.0\\
          & & &(1.64)&(97.2)\\
\colrule
Exper.\cite{chen} &         &  14.04   &      \\
\colrule

          &Nonmag   & 18.60 & 1.31 & 110.6\\
Struc-2b   &FM       & 19.04 & 1.59 & 99.5\\
          &Neel-AFM & 18.97 & 1.39 & 108.1\\
          &Coll-AFM & 18.84 & 1.40 & 108.8\\
          &Bicoll-AFM & -- & --& -- \\
\colrule
Exper.\cite{chenxiaolong1} &         &  20.48   &      \\
\colrule
          &Nonmag   & 15.46 & 1.30    & 113.9\\
Struc-4c   &FM       & 16.70 & 1.57    & 102.1\\
          &Neel-AFM & 15.75 & 1.35    & 111.2\\
          &Coll-AFM & 15.79 & 1.38    & 110.7 \\
          &Bicoll-AFM & 16.24&1.42&109.2\\
          & & &(1.64)&(96.7)\\
\colrule
Exper.\cite{chenxiaolong1} &         &  16.16   &      \\
\end{tabular}
\end{ruledtabular}
\end{table}

\begin{figure}
\includegraphics[width=8cm]{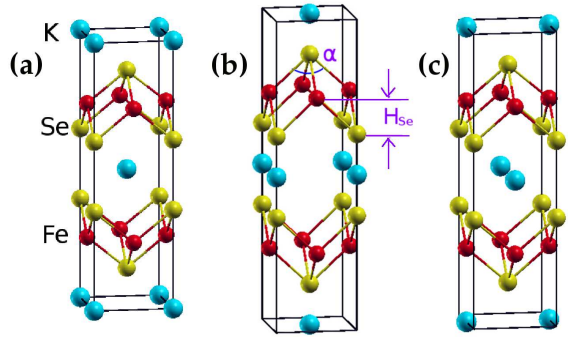}
\caption{(Color online) Schematic structure of
the tetragonal unit cell of
K$_x$Fe$_2$Se$_2$ containing two formula units:
 (a)K atom is situated in $2a$ Wyckoff position,
 (b)K atom is situated in $2b$ Wyckoff position,
 (c)K atom is situated in $4c$ Wyckoff position and only two of four $4c$ positions are occupied, the $\alpha$ and $H_{Se}$ indicate the Fe-Se-Fe bond angle and the Se height to Fe plane.
 } \label{figa}
\end{figure}

\section{Results and discussion}
%%################################################Structure############################################################################
For the first part, we study the structural properties of KFe$_2$Se$_2$ including the lattice parameter, the Se height to Fe plane $H_{Se}$ and the Fe-Se-Fe bond angle $\alpha$ in FeSe layer.
%% 1, discrible the calculated the structure%%
Alkali metal intercalated iron selenides have the tetragonal crystal structure with the space group I4/mmm. Fe atom and Se atom are sited in the $4d$ and $4e$ Wyckoff position in this space group respectively, constructing the stable FeSe layer in KFe$_2$Se$_2$ compound. In the previous studies KFe$_2$Se$_2$ is thought to have the ThCr$_2$Si$_2$-type structure and K atom is located at the $2a$ Wyckoff position similar to Ba atom in BaFe$_2$As$_2$, shown in Fig. \ref{figa}(a). The $2a$ position is at the body center of the Se tetragonal formed by eight Se atoms between two neighbor Fe layers, while the $2b$ and $4c$ positions are at the middle point of the side edge and the center of side face in the Se tetragonal body, see Fig. \ref{figa}(b) and (c).
%%(why did I consider the 2a 2b 4c position ?)%%

%Based the above idea,
We perform the structure optimizations of KFe$_2$Se$_2$ in which the K atoms is sited in $2a$, $2b$ and $4c$ Wyckoff position in space group I4/mmm respectively. The calculated results are showed in the Table. \ref{tab:table1} and the optimized lattice parameters $c$ are compared with the measured ones. When K atom take up the $2a$ position, the relaxed lattice constant $c$ is about 13.3 $\sim$ 14.5 \AA{} for different magnetic phases, consistent with the measured data 14.04 \AA{} synthesized by high-temperature routes. In the case of K occupying the $2b$ and $4c$ position, the optimized parameters $c$ are about 18.6 $\sim$ 19.0 \AA and 15.5 $\sim$ 16.7 \AA{} respectively for different magnetic orderings, with good agreement with the observed lattice $c$ 20.48 \AA{} and 16.16 \AA{} synthesized using the ammonothermal method.
%Even though the 18.6 $\sim$ 19.0 \AA{} is a little larger than 17.53 \AA{}, we think that the experimental lattice 20.48 \AA{} should be result from the mixed state of the $2b$ and $4c$ cases since it is easy for the Struc-$2b$(?) to be transformed to Struc-$4c$  in compound KFe$_2$Se$_2$ discussed in the section below.

%%describe the angle and Se height relationship with the Tc
The T$_c$ dependence of the As height to Fe plane and the Fe-As-Fe angle have been investigated extensively in iron arsenide superconductor and the neutron diffraction measurements reveal that T$_c$ becomes maximum when the Fe-As-Fe angle is close to 109.4 degree, corresponding to the perfect FeAs$_4$ tetrahedron \cite{zhaoj}.
The calculated Se heights to Fe plane $H_{Se}$ and the Fe-As-Fe angles $\alpha$ in three structure phases of KFe$_2$Se$_2$ are presented in Table. \ref{tab:table1}. In general, the Se heights within the given anti-ferromagnetic orders for three structure phases are in the range of 1.35 $\sim$ 1.45 \AA{} and the the discrepancy of Fe-As-Fe angel between our calculation and the ideal value 109.4$^\circ$ is less than 2.4$^{\circ}$ (except the angles along ferromagnetic direction of Bicoll-AFM order in Struc-2a and Struc-4c). Since the Coll-AFM magnetic phase is the lowest or second lowest energy state for three structure phases of KFe$_2$Se$_2$ (see Table.\ref{tab:table2}), it is convenient for us to compare the Fe-Se-Fe angles $\alpha$ in KFe$_2$Se$_2$ in this magnetic order. In our calculations
the angles in Struc-2a, Struc-2b and Struc-4c phases are 107.0$^\circ$, 108.8$^\circ$, and 110.7$^\circ$.
Among three structures, the 108.8$^\circ$ in Struc-2b is the most close to 109.4$^\circ$. According to the qualitative conclusion that the ideal tetrahedral FeAs$_4$ or FeSe$_4$ correspond to the maximum of T$_c$, we suggest that Struc-2b is related to the T$_c$=40 K, while Struc-2a and Struc-4c correspond to T$_c$ about 30 K superconduting phases.
As for the Se height to Fe plane, it is dependent on both the Fe-Se-Fe angle and the crystal lattice constants $a$ (or $b$), there is no definite correlation between Se height and T$_c$.

\begin{figure}
\includegraphics[width=7cm]{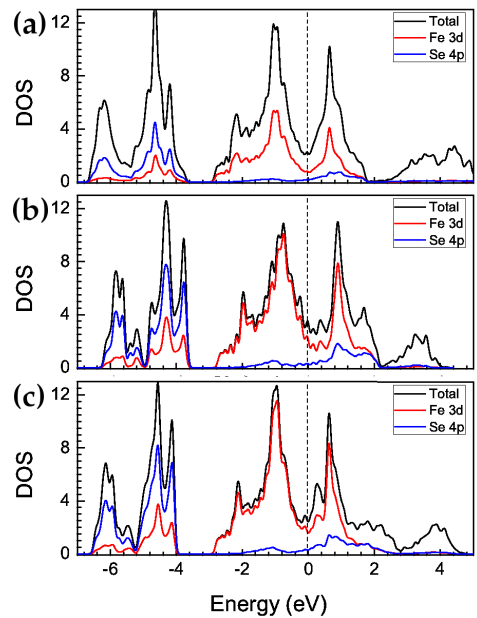}
\caption{(Color online)
Total and orbital-resolved partial density of states per primitive cell of KFe$_2$Se$_2$ for the Struc-2a, Struc-2b and Struc-4c structures. The Fermi level is set to zero.
 } \label{fig-dos}
\end{figure}

%%%discribe the nonmage electroonic structue , band fs and dos
For the second part, we investigate the electronic properties of the Struc-2a, Struc-2b and Struc-4c structures in nonmagnetic state.
%%%describle the dos
Fig. \ref{fig-dos} (a), (b) and (c) show the total and orbital resolved partial density of states (DOS) for three structures respectively. The DOS ranging from -7 to -3 eV are formed by the hybridization between Se 4p and Fe 3d orbitals. The DOS at Fermi level is mainly from Fe 3d orbital and a small contribution is from Se 4p orbital. The Struc-2b has the maximum 3.37 states/eV per unit cell, compared with 2.15 of Struc-2a and 2.43 of Struc-2c, which reveal that Struc-2b should be possessed of some especial electronic properties different from the Struc-2a and Struc-4c phase.

\begin{figure}
\includegraphics[width=8cm]{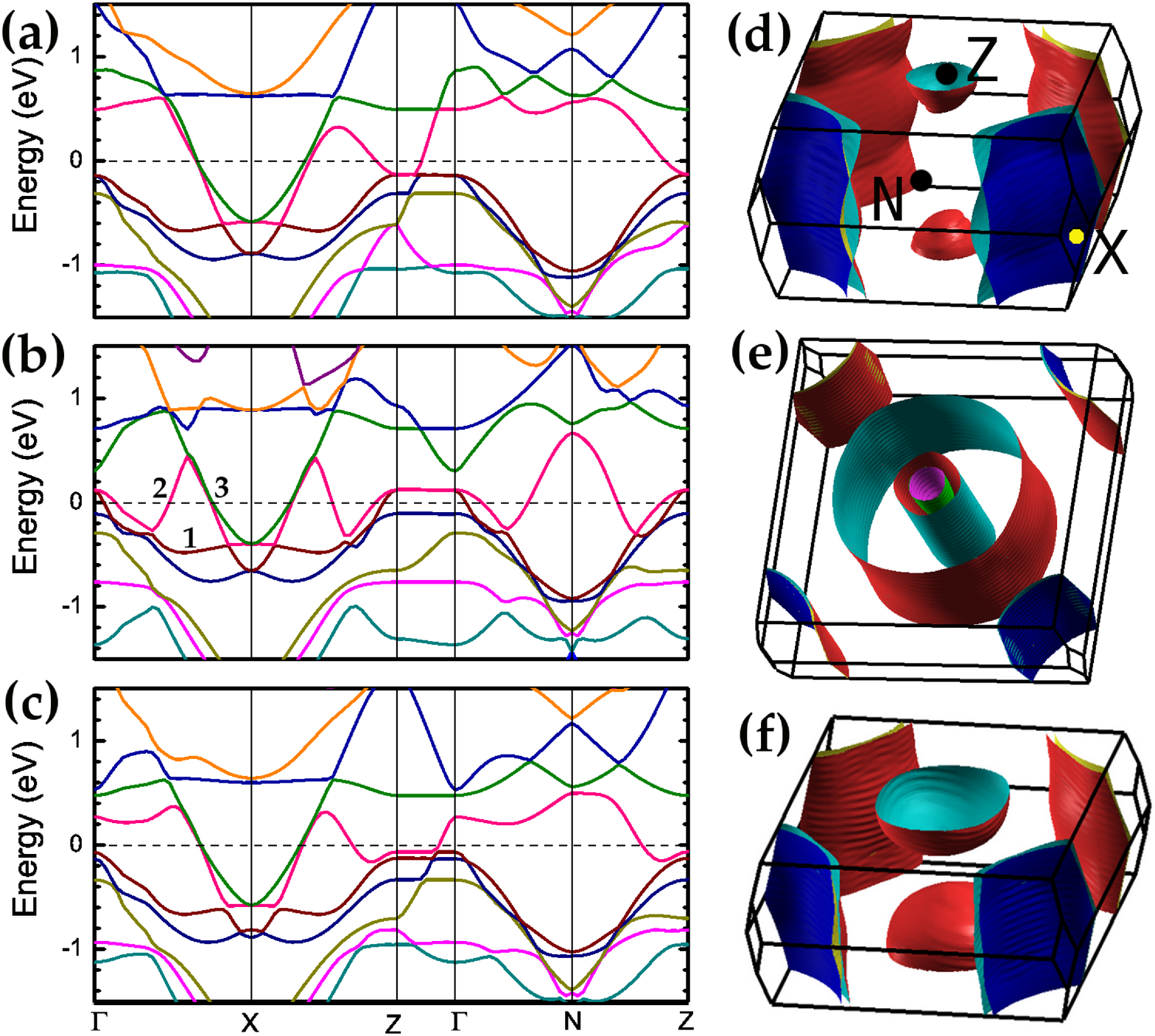}
\caption{(Color online)
The energy band structures of KFe$_2$Se$_2$ for the Struc-2a, Struc-2b and Struc-4c structure phase shown in (a), (b) and (c), and the corresponding Fermi surfaces shown in (d), (e) and (f) respectively. The bands crossing the Fermi energy denoted by line 1, 2 and 3 in (b). The Fermi level is set to zero.
 } \label{fs-band-1}
\end{figure}

Fig.\ref{fs-band-1} (a), (d) and (c), (f) show the band structures and Fermi surfaces of the Struc-2a and Struc-4c phase respectively. The two structure phases have a great resemblance in the band structure and the topological character of Fermi surface, including that two Fermi surface sheets at the Brillouin Zone corner and a Fermi surface pocket around Z point. They also share the critical feature that no Fermi surface appear near $\Gamma$ point obviously different from the iron arsenide superconductor. According to this resemblance, the Struc-2a and Struc-4c structure should have the similar superconducting behaviors and the T$_c$ should be approximate to each other, associated with the facts that T$_c$ are about 30 K in both the $c$=16.16 \AA{} and the $c$=14.04 \AA{} superconducting phases in KFe$_2$Se$_2$ compound.

From Fig.\ref{fs-band-1} (b), (e) we can see that there are three bands crossing the Fermi level for the Struc-2b and the Fermi surfaces present better 2-dimensionality characters than that of Struc-2a phase. The most surprising thing is that there exist three cylinder shape Fermi surfaces centered around $\Gamma$-Z, the innermost is hole-type derived from the band 1 marked in Fig.\ref{fs-band-1} (b) and the other two is electron-type derived from the energy band 2. The concentric cylinder Fermi surface sheets along $\Gamma$-Z is the typical features in iron arsenide superconductor, so the electronic structure of the Struc-2b phase of KFe$_2$Se$_2$ is very similar to the electronic structure of iron arsenide superconductor such as LaFeAsO and BaFe$_2$As$_2$. In Struc-2b phase, it is the emergence of cylinder Fermi surface around $\Gamma$-Z that is expected to bring about the stronger superconducting electron pairing than that in the Struc-2a and Struc-4c phases. The experimental T$_c$= 40 K superconducting phase should correspond to the Struc-2b phase in KFe$_2$Se$_2$ compound.

%%%%discribe the magnetic interaction
For the third part, we discuss the magnetic interaction between Fe moments and its connection to the superconducting temperature T$_c$.
The magnetic moments around each Fe atom were found to be about 2.8 $\mu B$ in the alkali metal intercalated iron selenide superconductors \cite{bao1}, much larger than the moments in the iron arsenide superconducting materials, so the magnetic interactions among Fe atoms play an important role in determination of structural and electronic properties of AFe$_2$Se$_2$ (A=K, Rb, Cs and Tl).
We perform the calculation for the Struc-2a, Struc-2b and Struc-4c structures under the nonmagnetic order (NM), ferromagnetic order (FM), checkerboard antiferromagnetic order (Neel-AFM), collinear antiferromagnetic order (Coll-AFM), and bi-collinear antiferromagnetic order (Bicoll-AFM). The magnetic ground states for the Struc-2a and Struc-4c are the Bicoll-AFM order. The lowest energy state for the Struc-2b is the Coll-AFM order, since under the Bicoll-AFM order K atoms will migrate to $4c$ position and lead to the Struc-4c structure in our calculations. The energies of three structures under several magnetic orders relative to the nonmagnetic energy of the Struc-2a are listed in the Table. \ref{tab:table2}.
\begin{table} [b]
\caption{\label{tab:table2}
The energy per Fe atom, magnetic moment around Fe atom and the nearest and next nearest neighbor superexchange coupling of KFe$_2$Se$_2$ system in different magnetic orderings for the Struc-2a, Struc-2b and Struc-4c structures. The nonmagnetic energy of Struc-2a is set to zero. For Struc-2b, the Bicoll-AFM state is not stable.
}
\begin{ruledtabular}
\begin{tabular}{clccc}
%\textrm{Left\footnote{Note a.}}&
%\textrm{Centered\footnote{Note b.}}&
%\multicolumn{1}{c}{\textrm{Decimal}}&
%\textrm{Right}\\
         &magnetic &Energy&moment&J$_1$,J$_2$\\
structure&phase    &(meV/Fe)&($\mu$B)&(meV/S$^2$)\\
\colrule
          &Nonmag   &   0.0   & 0    &  \\
Struc-2a   &FM       &-205.48  & 3.20 & J$_1$=-25.27 \\
          &Neel-AFM &-104.42 & 2.25 & J$_2$= 13.08\\
          &Coll-AFM &-207.41 & 2.75 &   \\
          &Bicoll-AFM &-248.25 & 2.75 &   \\
\colrule
          &Nonmag   & {}512.97 & 0     & \\
Struc-2b   &FM       & {}388.02 & 3.10 & J$_1$= 2.79\\
          &Neel-AFM & {}376.85 & 2.36  & J$_2$=20.75\\
          &Coll-AFM & {}298.32 & 2.85   & \\
          &Bicoll-AFM & -- & --   & \\
\colrule
          &Nonmag   & {}307.83 & 0.0  & \\
Struc-4c   &FM       &  { }77.39 & 3.16 & J$_1$=-24.48\\
          &Neel-AFM & {}175.31 & 2.37 & J$_2$= 11.86\\
          &Coll-AFM &  { }78.90 & 2.79 &  \\
          &Bicoll-AFM & { }30.77 & 2.79 &  \\
\end{tabular}
\end{ruledtabular}
\end{table}

To qualify the magnetic interactions, we can assume that the energy differences among these magnetic orders mainly originate from the interactions between each two Fe moments with spin $\vec{S}$. Then the frustrated Heisenberg model with the nearest , next nearest and third next nearest neighbor couplings $J_1$, $J_2$ and $J_3$ can be used to describe the interactions in KFe$_2$Se$_2$ system, namely,
\begin{equation}\label{eq:Heisenberg}
H=J_1\sum_{\langle ij \rangle}\vec{S}_i\cdot\vec{S}_j +J_2\sum_{ \ll ij \gg}\vec{S}_i\cdot\vec{S}_j +J_3\sum_{ \langle\langle\langle ij \rangle\rangle\rangle}\vec{S}_i\cdot\vec{S}_j,
 \end{equation}
whereas $\langle ij \rangle$, $\ll ij \gg$ and $\langle\langle\langle ij \rangle\rangle\rangle$ denote the summation over the nearest, next-nearest, and next-next-nearest neighbors, respectively. From the energy data in Table. \ref{tab:table2}, the superexchange coupling parameters J$_1$ and J$_2$ are shown in the rightmost column in Table. \ref{tab:table2}(The detailed calculation is referred to Appendix of Ref. \onlinecite{ma1}).

By comparing these exchange coupling values, we notice that the Struc-2a and Struc-4c phases have the similar features in the magnetic interactions, but the Struc-2b phase is somewhat different from them.
Here we should pay close attention to the next neighbor superexchange interaction $J_2$ in three structures.
Theoretically, the two band model $t-J_1-J_2$ indicates that the T$_c$ in iron based superconductor is proportional to the $J_2$ \cite{hujiangping} and we also have drawn the same conclusion in the previous first principles studies on the iron based materials \cite {ma-front,yan-caclfep}. From this point of view, we can obtain the results that Struc-2b phase has the higher superconducting transition temperature T$_c$ than Struc-2a and Struc-4c, and the latter two structures have the similar T$c$ values, because the next neighbor superexchange interaction $J_2=20.74~meV/S^2$ in Struc-2b is larger than $J_2=13.08~meV/S^2$ in Struc-2a and $J_2=11.86~meV/S^2$ in Struc-4c.
The opinion that the higher T$_c$ superconduting phase occur in the Struc-2b structure is consistent with our analysis about both the Fermi surface features and the Fe-Se-Fe angle.
The consistency is easy to understand because the superexchange interaction $J_2$ between two next nearest neighbor Fe atoms is bridged by Se atom and has a direct connection to the Fe-Se-Fe angle.

In addition, we do some discussion about the issue why K atom can locate in $2b$ or $4c$ Wyckoff position. Fixed the lattice parameters to the experimental values, we optimize the K positions with initial K position sited in the $2a$, $2b$ or $4c$ respectively. The K positions do not change from the initial $2a$ or $4c$ to $2b$ position with the fixed $c$ axis at 20.48 \AA, or from the initial $2a$ or $2b$ to $4c$ with the fixed $c$ axis at 16.16 \AA. These calculations indicate that the Struc-2b and Struc-4c are the metastable structure phases relative to the Struc-2a. As to the formation of these metastable phases, we can assume the process happened as blow. At first, the NH$_3$ molecules and K atoms enter in between FeSe layers together and K atoms sit in $2b$ or $4c$ Wyckoff positions, leading to the large length of lattice c. Then, NH$_3$ escape from the compounds and K atoms still sit in $2b$ or $4c$ position. We also investigate the energy change with the lattice parameter $c$ increasing for the Struc-2a, Struc-2b and Struc-4c respectively. When the length of lattice parameter $c$ is large than 18.58 \AA, the energy of the Struc-4c begin to become less than the one of Struc-2a phase. When the parameter $c$ is larger than 20.85 \AA, the Struc-2b energy is the lowest one among three structure phases. The results mean that with the large lattice parameter $c$ the K atom prefers to occupy the $4c$ or $2b$ Wyckoff position.

In summary, we have investigated the effect of different Wyckoff position of K atom on the crystal structure of KFe$_2$Se$_2$ by the first principle calculations. We conclude that the occupations of Wyckoff position $2b$ and $4c$ by K atom correspond to the T$_c$=40 K and T$_c$=30 K superconducting phases in experiment respectively, which are demonstrated by the calculated lattice parameters, Fe-Se-Fe angle, electronic structures and superexchange coupling $J_2$. Our results have a good agreement with the experiment and provide a reasonable approach to understand the new superconducting phases in KFe$_2$Se$_2$.

\section{Acknowledgments}
XWY sincerely thanks Prof. Shaojing Qin for giving the opportunity of visiting ITP, during the visiting period the manuscript was finalized. This work is partially supported by National Program for Basic Research of MOST (2011CBA00112) and by National Natural Science Foundation of China (U1204108).

\end{document}